\documentclass[12pt]{iopart}

\usepackage{graphics,epsfig,subfigure}
\usepackage{iopams}

\newcommand{\bin}[2]{C^{#2}_{#1}}
\newcommand{\nn}{\nonumber}
\newcommand{\bb}[0]{\begin{eqnarray}}
\newcommand{\ee}[0]{\end{eqnarray}}

\renewcommand{\eref}[1]{Equation~(\ref{#1})}
\newcommand{\efig}[1]{Figure~\ref{#1}}
\newcommand{\cst}{{\rm cst}}

\newcommand{\ff}{\frac{1}{2}}

\renewcommand{\br}{{\bf r}}
\newcommand{\bq}{{\bf q}}
\newcommand{\ve}{{\bf e}}

\newcommand{\erf}{{\rm erf}}
\newcommand{\tg}{{\tilde g}}
\newcommand{\tk}{{\tilde \kappa}}
\renewcommand{\e}{{\rm e}}
\newcommand{\co}{\beta}

\begin{document}
\title{Nature of the global fluctuations in the spherical model at
criticality}

\author{Jean-Yves Fortin and Sophie Mantelli}
\address{CNRS, Institut Jean Lamour, D\'epartement de Physique de la Mati\`ere
et des Mat\'eriaux,
UMR 7198, Vandoeuvre-les-Nancy, F-54506, France\\
}

\ead{fortin@ijl.nancy-universite.fr,smantell@ijl.nancy-universite.fr}
\date{\today}
\begin{abstract}
We study the universal nature of global fluctuations in the critical regime of
the spherical model by evaluating the exact distribution of the magnetization
and its absolute value in the thermodynamical limit, in the presence of
a conjugate field. We show that the probability distribution function for this
model is described by non-Gaussian asymptotics and non-symmetric characteristics
which depend on the dimension of the system $2<d<4$.
Relation with extreme statistics of independent wavelength modes is discussed.
\end{abstract}


\maketitle

\section{Introduction}

Global fluctuations of space-averaged order parameters in many statistical
systems possess generally a non-gaussian behavior in their critical and non-mean
field regime. One well-known example is given by the two-dimensional XY-model
(2D-XY), whose magnetic fluctuations (specifically the absolute value of the
global spin vector)
can be studied within the spin-wave approximation on
the low-temperature critical line~\cite{arch98,prl00,pre01}, far enough from
the Kosterlitz-Thouless transition~\cite{berezinskii71,KT73,Nelson77}. One
interesting feature and universal character of the resulting rescaled
probability distribution function (PDF) is that it does not depend on the
critical exponents,
temperature and system size~\cite{bhp}. Moreover the exponential and
double exponential asymptotic fall-off behaviors for negative and positive
deviations respectively, up to some corrective terms,
identify this PDF to a general Gumbel function parametrized by a non-integer.
Gumbel functions describe the extrema distribution of a set of independent and identically distributed Gaussian variables.
They are parametrized by an integer which is equal to one for the lowest
value, two for the second lowest values and so on.
The connection with the 2D-XY-model is based on the existence of low excitations
modes, the Goldstone modes, which tend to destroy the quasi-long range order at
low temperature. In Fourier space, the modes of value $\bq$, are decoupled and
have a mass linearly proportional to $|\bq |$, and therefore the low soft modes
are the predominant contributions to the PDF, giving rise to a strongly
non-Gaussian behavior and
Gumbel limiting distribution with non-integer parameter $\pi/2$. Simple one
dimensional and self-critical models such as interface distribution or $1/f$
noise~\cite{antal01} give also a Gumbel distribution with parameter equal to one
and therefore strongly non-Gaussian.
Gumbel distributions with generalized parameter are also common in one-dimensional non-equilibrium and correlated systems
where the distribution can be cast into a product of single probabilities with
independent but non-identically distributed variables~\cite{bertin05,Bertin06}.
In all
these example, the Fourier modes have mass proportional to $|\bq |$, therefore
leading to the same limiting distribution dominated by the soft modes.
This function is constructed by isolating the non-singular part of the
fluctuations when the
system size is taken to infinity, usually this is done by rescaling the order
parameter using the mean value and variance only.
 Self-organized systems~\cite{bak87,jensen} and network applications are also
important systems in the way that they often lead to exact mathematical
results~\cite{seneta69,biggins93}, for example the
distribution of connectivity in random graphs~\cite{Barabasi99}, width
distribution of interfaces~\cite{racz94}, or size distributions in replicative
phenomena~\cite{JSKim07,Jo10}.
Generally, one expects to approach the critical point from the low
temperature region where the distribution for the order parameter $m$
(for example the magnetization per site in a spin system) can
be generally written as $P(m,L)=L^{\beta/\nu}Q(m
L^{\beta/\nu},\xi/L,h L^{\beta\delta/\nu})$, where $\xi$ is the correlation
length, $h$ an external field conjugated to the order parameter $m$, $L$ the
typical
system size and $Q$ the limiting or asymptotic PDF independent of $L$.
The critical exponents $(\beta,\delta,\nu)$ depend on the specific model.
For the classical 2D-XY model, where $m$ is defined as the modulus of the total
spin vector divided by the number of sites $L^2$, this asymptotic distribution
can be very closely written as~\cite{pre01}

\bb
Q(\theta)\simeq \exp\Big (-\frac{1}{8\pi}\e^{8\pi\Big (\sqrt{g_2/2}\theta-a_0
\Big )}+8\pi \sqrt{\frac{g_2}{2}}\theta \Big )
\ee

with $a_0=1/24+\gamma/(4\pi)-\ln(4\pi)/(4\pi)-\ln \prod_{k=1}^{\infty}\Big
[1-\exp(-2\pi k)\Big ]/(2\pi)$ $\simeq 0.113\,514$, and constant $g_2\simeq
3.867\times 10^{-3}$. Here $\theta=(m-<m>)/\sigma$, with $\sigma^2$ the
variance, is the standard rescaled parameter for defining correctly the
non-singular limit.
The asymptotic regimes when $|\theta|$ is large are important in defining the
class of function where the PDF belongs to. For independent and identically
distributed fluctuations, the PDF falls into the Gaussian class. Otherwise,
the PDF usually falls into class of functions characteristic of generalized
extreme value statistics, for example Gumbel, Weibull or
Fr\'echet~\cite{bertcl08}. They are however few correlated systems where
the asymptotic PDF of global order parameter can be evaluated
analytically~\cite{plischke94,pre01} and where the effect of the correlations
can be measured. In this paper
we focus essentially on the critical regime of a simple correlated spin-model,
the well-known spherical model in dimension $2<d<4$~\cite{berlin52}, where
fluctuations of the order parameter (magnetization and its absolute value)
can be studied exactly in the critical regime. This will give some
physical insight on universal features belonging to critical correlated systems.

\section{Definition of the Model}

We consider the spherical model~\cite{berlin52} defined by a set of $N=L^d$
scalar spins $-\infty< S_i<+\infty$ coupled together with nearest-neighbor
ferromagnetic coupling $J_{ij}$ on a $d$-dimensional lattice at temperature
$T=1/\beta$. The spins are constraint by the condition $\sum_i S_i^2=N$, and the
Hamiltonian is given by

\bb
H_0=-\frac{1}{2}\sum_{i,j}J_{ij}S_iS_j,
\ee

with $J_{ij}=J\sum_{\alpha}\delta_{\br_i,\br_j\pm \ve_{\alpha}}$ where $\ve_{\alpha=1,\cdots,d}$ are the unit
spacing vectors of the lattice. For example, in $d=2$, $\ve_1={\bf x}$ and $\ve_2={\bf y}$. Following the standard
techniques, see~\cite{berlin52,baxter}, the partition function is written using
a Dirac delta function representation to impose the constraint on the spins

\bb\fl
Z_N(h)=\int \frac{ds}{2\pi}\int_{-\infty}^{+\infty}\prod_{\br_i}dS_i\exp\left [ -\beta H_0+h\sum_i S_i  +(is+a)\left (
N-\sum_iS_i^2\right )\right ]
\ee


\begin{figure}
\centering
\includegraphics[scale=0.5]{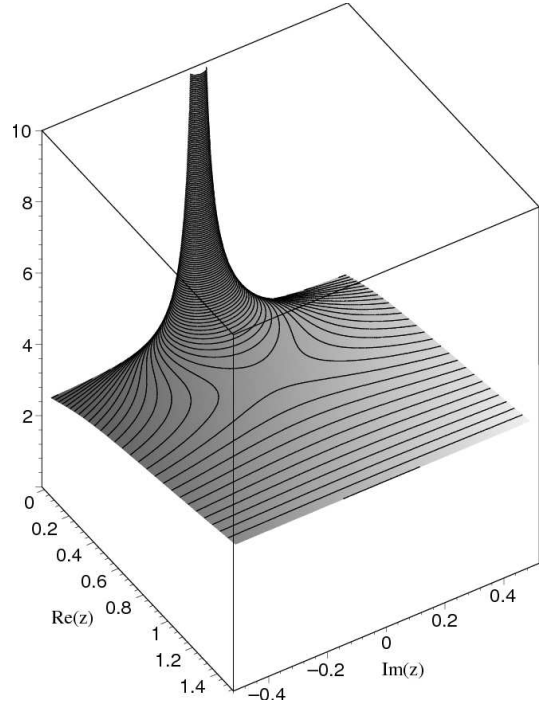}
\caption{\label{phi} Surface plot of the function $\Phi(z,h)$ (vertical axis) in
the complex
plane, showing the location of the unique saddle point. Here $h=K=1$ and
$N=10^{\,3}$ in $d=3$.}
\end{figure}

where $a$ is any arbitrary real scalar which does not change the value of the integral over $s$.
The successive derivatives of $Z_N(h)$ with respect to $h$ give the moments of
the magnetization $m:=\sum_iS_i/N$ which we want to study.
We will also consider another partition function by replacing the field term
$h\sum_i S_i$ by $\bar h\left
|\sum_i S_i \right |$, and from which we can derive the successive moments of
the absolute
value of the magnetization $|m|$.
In general we can consider both fields and study the different moments from $Z_N(h,\bar h)$. The integration over the
$S_i$ can be done using Fourier transform $S_i=\frac{1}{\sqrt{N}}\sum_{\bq}S_{\bq}e^{i\bq.\br_i}$,
with $S_{-\bq}=\bar S_{\bq}$, $\bq=2\pi(k_1,\cdots,k_d)/L$, and $k_i=0,\cdots,L-1$. In this case the interaction term can be diagonalized directly such as

\bb\nn
\frac{\beta}{2}\sum_{i,j}J_{ij}S_iS_j
=K\sum_{\bq}S_{\bq}S_{-\bq}\sum_{\alpha}\cos\,q_{\alpha}
\ee

with $K:=\beta J$. After Fourier transformation, the constraint contribution
becomes simply

\bb\fl
(is+a)\sum_iS_i^2=\frac{(is+a)}{N}\sum_i\sum_{\bq,\bq'}S_{\bq}S_{\bq'}e^{i(\bq+\bq').\br_i}
=(is+a)\sum_{\bq}S_{\bq}S_{-\bq}.
\ee

The other linear terms are equal to $h\sum_i S_i = h\sqrt{N}S_0$, and $\bar
h\left |\sum_i S_i\right | = \bar h\sqrt{N}\left |S_0\right |$.
We can finally express the partition function as an integral over a product of
decoupled Fourier modes

\bb\nn\fl
Z_N(h,\bar h)&=&\int \frac{ds}{2\pi}\int_{-\infty}^{+\infty}dS_0\prod_{\bq\ne 0}^{\hspace{0.8cm}'}dS_{\bq}dS_{-\bq}
\exp\left [ -K\sum_{\bq}S_{\bq}S_{-\bq}\left
(z+d-\sum_{\alpha}\cos\,q_{\alpha}\right )\right .
\\ \label{ZNq}\fl
&+&\left .h\sqrt{N}S_0+\bar h\sqrt{N}\left |S_0\right |+NK(z+d)\right ]
\ee

where $K(z+d)=is+a$. $a$ is chosen to avoid the integral divergence for the zero mode. The prime symbol on the product means
that only half of the Fourier modes are considered, using the symmetry
$\bq\rightarrow -\bq$. Decomposing $S_{\bq}=x_{\bq}+iy_{\bq}$ in real and
imaginary parts, so that $dS_{\bq}dS_{-\bq}=dx_{\bq}dy_{\bq}$,
$S_{\bq}S_{-\bq}=x_{\bq}^2+y_{\bq}^2$, we obtain after integration over
Fourier modes the integral expression of the partition function

\bb\nn\fl
Z_N(h,\bar h)&=&\int \frac{ds}{2\pi}\left ( \frac{\pi}{2K}\right )^{N/2}
\exp\left [ -\frac{1}{2}\sum_{\bq}\ln\left (z+d-\sum_{\alpha}\cos\,q_{\alpha}\right )+N\frac{h^2+\bar h^2}{4Kz}
+NK(z+d)\right ]
\\ \nn \fl
&\times&\frac{1}{\sqrt{2}}\left [ 2\cosh\frac{h\bar h}{2Kz}+e^{-Nh\bar h/(2Kz)}\erf\left (\frac{\sqrt{N}(\bar h-h)}{2\sqrt{Kz}}\right )
+e^{Nh\bar h/(2Kz)}\erf\left (\frac{\sqrt{N}(\bar h+h)}{2\sqrt{Kz}}\right ) \right ].
\ee

From this expression, we can define the intensive free energy $\Phi(z,h,\bar h)$
for the system such that

\bb
Z_N(h,\bar h)&=&\int \frac{ds}{2\pi}\left ( \frac{\pi}{2K}\right )^{N/2}{\sqrt{2}}
\exp\left [ N\Phi(z,h,\bar h) \right ]
\ee

with, for each of the two order parameters considered in this paper

\bb\label{phidef}\fl
\Phi(z,h,0)&=&-\frac{1}{2N}\sum_{\bq}\ln\left (z+d-\sum_{\alpha}\cos\,q_{\alpha}\right )+\frac{h^2}{4Kz}
+K(z+d),
\\ \fl\nn
\Phi(z,0,\bar h)&=&-\frac{1}{2N}\sum_{\bq}\ln\left (z+d-\sum_{\alpha}\cos\,q_{\alpha}\right )+\frac{\bar h^2}{4Kz}
+K(z+d)+\frac{1}{N}\ln\left [1+\erf\left (\frac{\sqrt{N}\bar h}{2\sqrt{Kz}}\right )\right ].
\ee

It is clear that $\Phi$ is not an even function of $\bar h$ for the absolute
value of magnetization.
As $N$ is large, we can study the saddle points of $\Phi$ in the complex plane,
and obtain the main contributions to the free energy and partition function.

\section{Saddle point analysis and scaling relations in the critical region}

In  \efig{phi}, we represented $\Phi$ for some arbitrary values of parameters, which shows the location of the unique saddle point as the minimum of $\Phi$ along the real $z$-axis, which is also a maximum along the real $s$-axis or imaginary $z$-axis. In particular
the equation of the local saddle point satisfies for example

\bb\label{saddle}
\frac{\partial\Phi(z,h,0)}{\partial z}=-\frac{1}{2N}\sum_{\bq\neq 0}\frac{1}{z+d-\sum_{\alpha}\cos\,q_{\alpha}}-\frac{1}{2Nz}
-\frac{h^2}{4Kz^2}+K=0.
\ee

The last equation gives the implicit saddle point solution $z(h)$ as function of
field $h$. The critical point $K=K_c$ is defined by
$K_c:=\lim_{N\rightarrow \infty}\frac{1}{2N}\sum_{\bq\neq
0}(d-\sum_{\alpha}\cos\,q_{\alpha})^{-1}=:\ff g_1$, after taking first the limit
$N\rightarrow\infty$ then $h=0$ and finally $z=0$. This quantity is well defined
for $d>2$. In two dimensions for example, this quantity is pertinent for
characterizing the XY model and diverges with the system size as $g_1\simeq
(1/2\pi) \log (CN)$~\cite{pre01} with the exact value

\bb\nn
C=\exp\Big \{\frac{\pi}{3}+2\log \Big (\frac{\sqrt{2}}{\pi}\Big )+2\gamma-4\log \prod_{n=1}^{\infty}[1-\exp(-2\pi n)]\Big \}\simeq 1.8456
\ee

The evaluation of this constant requires a precise analysis of the discrete sum
over the Fourier modes which is specific to the geometry of the lattice. By
extension it is useful to define the
dimension-dependent quantities

\bb
g_p:=
\frac{1}{N^{p}}\sum_{\bq\neq 0}\frac{1}{\left (
d-\sum_{\alpha}\cos\,q_{\alpha}\right )^p}=:\frac{1}{N^p}\sum_{\bq\neq
0}G_{\bq}^p
\ee

which scale with $L$ in the large size limit, and where $G_{\bq}$ is the Green
function. Indeed, in the interval $2<d<4$, we can replace the sum over modes
$\bq$ by an integral and show that

\bb
g_p
\approx \frac{1}{N^{p}}
\int_{\cst/L}^{q_{max}}\frac{L^d}{(2\pi)^d} G_{\bq}^pq^{d-1}dq,
\ee

where $q_{max}$ is some cut-off dependent of the lattice step. The integrand is
dominant when $q$ is small, precisely $G_{\bq}\propto q^{-2}$, then $g_p\approx
L^{d(1-p)}[q^{d-2p}]_{\cst/L}^{q_{max}}$. If $p=1$, the integral is convergent
for
$d>2$, and $g_1$ is finite, as well as the critical temperature. Otherwise, for
$p\ge 2$,
$d-2p$ is strictly negative if $d<4$, and the integration part
$[q^{d-2p}]_{\cst/L}^{q_{max}}$ is dominated by
$L^{2p-d}$ which is divergent. Therefore $g_p$ scales like
$L^{d(1-p)+2p-d}=L^{-p(d-2)}$ and
goes to zero like

\bb
g_{p\ge 2}= \tg_p L^{-p(d-2)}+O(L^{-d(p-1)}),\;\tg_p:=\lim_{L\rightarrow \infty}\frac{1}{L^{2p}}\sum_{\bq\neq 0}G_{\bq}^p,
\ee

where $\tg_p$ are finite and positive constants. The corrective terms
$O(L^{-d(p-1)})$ are
always negligible
since $2p>d$, which holds when $2<d<4$. The case $d=2$ is particular since all $g_p$ are
finite, except for $g_1$ which diverges logarithmically with $L$ as it was
discussed above. These constants are actually essential
for evaluating the PDF of the 2D-XY model~\cite{pre01} as they are related to
the
moments of the distribution. For example, magnetization goes slowly to zero with
the system size like $<m>=(NC)^{-T/(8\pi)}=\e^{-T g_1/2}$.
The important point is that these constants depend only on dimension and
lattice geometry.

The scaling exponents of the spherical model in the range $2<d<4$ are given by
$\alpha=-(4-d)/(d-2)$, $\beta=1/2$, $\gamma=2/(d-2)$,
$\delta=(d+2)/(d-2)$, and $\nu=1/(d-2)$. They satisfy the scaling relations
$\alpha+2\beta+\gamma=2$, $\alpha=2-d\nu$, and $\gamma=\nu(2-\eta)$.
 In the critical region, we define therefore the rescaled parameters

\bb\label{scaling}
K-K_c=:k_0 L^{-1/\nu}=k_0L^{-(d-2)},\;h=:h_0L^{-\beta\delta/\nu}=h_0
L^{-(d+2)/2},
\ee

where $k_0$ (which can be negative, in the paramagnetic region) and $h_0$ are
external parameters. Similarly, $\bar h$ follows the same scaling as $h$, or
$\bar h=:\bar h_0 L^{-(d+2)/2}$. Therefore the
magnetization scales like $<m>\propto (K-K_c)^{\beta}\propto h^{1/\delta}\propto
L^{-(d-2)/2}$. These arguments imply that the different moments and cumulants of
order $p$ scale like $<m^p>\propto L^{-p(d-2)/2}$ in the critical region.
Cumulants $\kappa_p$ can be computed using the successive derivatives of $\Phi$:

\bb\label{kp}
\kappa_p=\frac{1}{N^{p-1}}\frac{d^p \Phi}{d\,h^p}=:\tk_p L^{-p(d-2)/2},
\ee

and the asymptotic distribution defined in the introduction can be identified
with the function $Q$ using a Fourier transform (see \ref{app1})

\bb
Q(\theta):=
\int\frac{d\lambda}{2\pi}\exp \left [ i\lambda\theta
+\sum_{p\geq 2}\frac{\tk_p}{\tk_2^{p/2}} \frac{(-i\lambda)^p}{p!}\right ],
\ee

which is independent of the system size. We first evaluate the saddle point
solution of
\eref{saddle}, using the scaling \eref{scaling} and assuming $z$ small:

\bb\nn
&-&\frac{1}{2N}\sum_{\bq\neq 0}\frac{1}{d-\sum_{\alpha}\cos\,q_{\alpha}}+
\frac{z}{2N}\sum_{\bq\neq 0}\frac{1}{\left (d-\sum_{\alpha}\cos\,q_{\alpha}\right )^2}+\cdots
-\frac{1}{2Nz}
\\
&-&\frac{h_0^2}{4K_cz^2}L^{-(d+2)}+
K_c+k_0L^{-(d-2)}+\dots=0.
\ee

A first analysis of the previous equation leads to the identification of the
scaling law $z\sim L^{-2}$ since for example $z^{-2}L^{-(d+2)}\sim L^{-(d-2)}$
by comparison of terms on the second line.
All other terms scale with the same exponent $2-d$. This means that $\partial \Phi/\partial z$ scales
like $L^{-(d-2)}$ and $\Phi\sim L^{-d}$ at the critical point.
Therefore the  solution $z(h)$ can be expanded as a series in the inverse power
of $L$: $z(h)=L^{-2}z_0+\cdots$ (and $z(\bar h)=L^{-2}\bar z_0+\cdots$ as well).
We obtain at the leading order $O(L^{-(d-2)})$

\bb\nn
&-&\ff g_1+\ff z_0\tg_2L^{-(d-2)}-\ff z_0^2\tg_3L^{-(d-2)}+\ff z_0^3\tg_4L^{-(d-2)}-\cdots
\\
&-&\frac{1}{2z_0}L^{-(d-2)}
-\frac{h_0^2}{4K_cz_0^2}L^{-(d-2)}+K_c+k_0L^{-(d-2)}=0.
\ee

The zero$^{th}$ order term $K_c=g_1/2$ is canceled, and we obtain an implicit
equation for $z_0=z_0(k_0,h_0)$ and $\bar z_0=\bar z_0(k_0,\bar h_0)$ as
function of the reduced temperature and magnetic field:

\bb
-\frac{1}{2z_0}+\ff z_0\tg_2-\ff z_0^2\tg_3+\ff
z_0^3\tg_4-\cdots=\frac{h_0^2}{4K_cz_0^2}-k_0,
\ee

and a similar expansion for $\bar z_0$ as well.
We can express the previous equation in a more compact form,
using the definition of the $\tg_p$s and summing the series over $p$

\bb\label{solz0}
-\frac{1}{2z_0}+\frac{1}{2}\sum_{\bq\neq 0}
\frac{z_0G^2_{\bq}L^{-4}}{1+z_0G_{\bq}L^{-2}}
=\frac{h_0^2}{4K_cz_0^2}-k_0.
\ee

For the conjugate field of the absolute value $|m|$, we obtain the same
expression with an additional term associated to the error function

\bb\label{solz0b}\fl
-\frac{1}{2\bar z_0}+\ff\sum_{\bq\neq 0}
\frac{\bar z_0G^2_{\bq}L^{-4}}{1+\bar z_0G_{\bq}L^{-2}}
=\frac{\bar h_0^2}{4K_c\bar z_0^2}-k_0+\frac{\bar h_0}{2\sqrt{\pi
K_c}\bar z_0^{3/2}}\frac{\e^{-\bar h_0^2/(4K_c\bar z_0^2)}}{1+\erf\Big
(\frac{\bar h_0}{2\sqrt{K_c \bar z_0}}\Big )}.
\ee

When $h_0=0$, or $\bar h_0=0$, and $k_0$ large, the previous equations give the
asymptotic solutions $z_0\sim \bar z_0 \sim 1/(2k_0)$. Numerically we solved
\eref{solz0} and \eref{solz0b} recursively to obtain the real solution $z_0$
and $\bar z_0$.

\section{Cumulant expansion and asymptotic distribution}

In this section we derive the cumulants from \eref{kp} and
given function $\Phi$, using the constraint
$\partial \Phi/\partial z=0$ of the saddle point imposed for any conjugate
field $h$ or $\bar h$. Indeed the identity

\bb
\frac{d^p(\partial_z \Phi(z(h),h,0))}{dh^p}=0
\ee

is valid for any integer $p\geq 0$, which simply means that at the saddle point
value $z=z(h)$, $\partial_z \Phi$ is always zero as function of $h$.
In
particular, we can show recursively in \ref{app2} that for $p\ge 2$

\bb\label{cumulant}
\kappa_p=\frac{1}{N^{p-1}}\frac{d^{p}\Phi}{dh^{p}}
=\frac{1}{N^{p-1}}\sum_{k=0}^{p}\bin{p}{k}\frac{\partial^p\Phi}{\partial z^k\partial h^{p-k}}
z'(h)^k.
\ee

Knowing the function $\Phi$, it is straightforward to obtain the cumulant
expression by derivation and summation. It is worth noting that cumulants
depend only on $z(h)$ and its first derivative.
From the scaling form of the distribution function, the scaling
law for the cumulants (or equivalently for the moments) is $\kappa_p\sim
L^{-p(d-2)/2}$.
 For example, $\kappa_1=<m>=h/(2Kz)=\frac{h_0}{2K_cz_0}L^{-(d-2)/2}=\tk_1
L^{-(d-2)/2}$.
We can generally use the fact that $dz(h)/dh=dz_0(h_0)/dh_0\times
L^{(d-2)/2}=:z_0'(h_0) \times L^{(d-2)/2}$, to
calculate the regular part $\tk_p$ of the cumulants near
the critical point, and obtain for example, using \eref{kappa},

\bb \label{kappa0}\fl
\frac{(-1)^p}{p!}\tk_p&=&\frac{z_0'^p}{2pz_0^p}+\frac{z_0'^{p-2}}{4K_cz_0^{p-1}}
+\frac{z_0'^p}{2p}\sum_{\bq\ne 0}\frac{1}{\left (z_0+L^2G^{-1}_{\bq} \right )^p}
-\frac{h_0z_0'^{p-1}}{2K_cz_0^{p}}+\frac{h_0^2z_0'^{p}}{4K_cz_0^{p+1}},
\ee

from which we deduce the second cumulant

\bb\label{kappa2}
\tk_2=\frac{z_0'^2}{2z_0^2}+\frac{1}{2K_cz_0}
+\frac{z_0'^2}{2}\sum_{\bq\ne 0}\frac{1}{\left (z_0+L^2G^{-1}_{\bq} \right )^2}
-\frac{h_0z_0'}{K_cz_0^{2}}+\frac{h_0^2z_0'^{2}}{4K_cz_0^{3}}
\ee

for order parameter $m$, and

\bb\label{kappa2b}
\tk_2=\frac{\bar z_0'^2}{2\bar z_0^2}+\frac{1}{2K_c\bar z_0}
+\frac{\bar z_0'^2}{2}\sum_{\bq\ne 0}\frac{1}{\left (\bar z_0+L^2G^{-1}_{\bq}
\right )^2}
-\frac{1}{\pi K_c\bar z_0}-\frac{\bar z_0'}{\sqrt{\pi K_c}\bar z_0^{3/2}}
\ee

for order parameter $|m|$, when $\bar h_0=0$ for
simplification. In the former case when $h_0=0$, we can show that $z'_0(0)=0$.
Indeed, deriving the saddle-point solution \eref{solz0}, we obtain

\bb\label{z0p}
z_0'\left (
\frac{h_0^2}{K_cz_0^3}+
\frac{1}{z_0^2}+ \sum_{\bq\neq 0}
\frac{G^2_{\bq}L^{-4}}{(1+z_0G_{\bq}L^{-2})^2}
\right )=\frac{h_0}{K_cz_0^2}
\ee

for conjugate field $h_0$ and

\bb\label{z0pb}
\bar z_0'\left (
\frac{1}{\bar z_0^2}+ \sum_{\bq\neq 0}
\frac{G^2_{\bq}L^{-4}}{(1+\bar z_0G_{\bq}L^{-2})^2}
\right )=\frac{1}{\sqrt{\pi K_c}\bar z_0^{3/2}}.
\ee

by deriving \eref{solz0b} and taking $\bar h_0=0$ afterward. The former equation
gives $z'_0(0)=0$, then only $\tk_2=1/(2K_cz_0)$ is non-zero in this
limit which implies that distribution $Q(\theta)$ is purely Gaussian with a
variance depending on an implicit equation for $z_0$.
However in the latter case \eref{z0pb}, $\bar z'_0(0)\ne 0$ instead and
$Q(\theta)$ is non-Gaussian since all the moments do not vanish.
We can in general derive formally $Q(\theta)$ from
definition \eref{Plim},
using the expression \eref{cumulant} for $\tk_p$, and the
characteristic function. We also redefine the free energy
$\Phi_0:=L^{d}\Phi(z_0L^{-2},h_0L^{-(d+2)/2})$, with

\bb\label{phi0}\fl
\Phi_0(z_0,h_0)=-\ff\log z_0+\ff\sum_{\bq\neq 0}
\left [
\frac{z_0G_{\bq}}{L^2}-\log\left (1+\frac{z_0G_{\bq}}{L^2}\right )
\right ]
+\frac{h_0^2}{4K_cz_0}+k_0z_0+\cst.
\ee

For the distribution of $|m|$ and conjugate field $\bar h_0$, we
find instead

\bb\nn\fl
\bar \Phi_0(\bar z_0,\bar h_0)&=&-\ff\log\bar z_0+\ff\sum_{\bq\neq 0}
\left [
\frac{\bar z_0G_{\bq}}{L^2}-\log\left (1+\frac{\bar z_0G_{\bq}}{L^2}\right )
\right ]
+\frac{\bar h_0^2}{4K_c\bar z_0}+k_0\bar z_0
\\ \label{phi0b}\fl
&+&
\log\left [
1+\erf\left (\frac{\bar h_0}{2\sqrt{K_c\bar z_0}}\right ) \right ]
+\cst.
\ee

The characteristic function $H$ appearing in the definition of $Q(\theta)$ in
\eref{charH} has the scaling form $\tilde H$ depending on $\Phi_0$ (or
$\bar \Phi_0$) and its derivatives

\bb
\tilde H(-\lambda)=\sum_{p\geq 2}\frac{(-i\lambda)^p}{\tk_2^{p/2}p!}\tk_p
=\sum_{p\geq 2}\frac{(-i\lambda)^p}{\tk_2^{p/2}p!}
\sum_{k=0}^{p}\bin{p}{k}\frac{\partial^p\Phi_0}{\partial z_0^k\partial h_0^{p-k}}
z_0'(h_0)^k.
\ee

The series over $p$ can be performed directly if we use for example the
Fourier transform of $\Phi_0$

\bb\nn
\Phi_0(z_0,h_0)&=&\int\int \frac{dqd\omega}{(2\pi)^2}\tilde \Phi_0(q,\omega)\exp(iqz_0+i\omega h_0),
\\
\tilde \Phi_0(q,\omega)&=&\int\int dz_0dh_0\Phi_0(z_0,h_0)\exp(-iqz_0-i\omega
h_0).
\ee

Indeed, we obtain formally the functional expression of the characteristic
function depending on $\Phi_0$ only

\bb\nn\fl
\sum_{p\geq 2}\frac{(-i\lambda)^p}{\tk_2^{p/2}p!}\tk_p
&=&\int\int \frac{dqd\omega}{(2\pi)^2}\tilde \Phi_0(q,\omega)\exp(iqz_0+i\omega h_0)
\sum_{p\geq 2}\frac{(-i\lambda)^p}{\tk_2^{p/2}p!}
\sum_{k=0}^{p}\bin{p}{k}(iq)^k(i\omega)^{p-k}
z_0'(h)^k
\\ \fl
&=&\int\int \frac{dqd\omega}{(2\pi)^2}\tilde \Phi_0(q,\omega)\exp(iqz_0+i\omega h_0)
\sum_{p\geq 2}\frac{(-i\lambda)^p}{\tk_2^{p/2}p!}
(iqz_0'+i\omega)^p
\\ \nn
&=&\Phi_0(z_0-i\frac{\lambda z_0'}{\sqrt{\tk_2}},h_0-i\frac{\lambda}{\sqrt{\tk_2}})
-\Phi_0(z_0,h_0)+i\frac{\lambda}{\sqrt{\tk_2}}\left (
z_0'\frac{\partial}{\partial z_0}+\frac{\partial}{\partial h_0}
\right )\Phi_0(z_0,h_0).
\ee

This relation satisfies the constraint $\tilde H(0)=0$ coming from the
normalization of the PDF. It is also straightforward from \eref{Plim} to obtain
the general form of the limiting distribution for conjugate
fields $h_0$ and $\bar h_0$ as well by replacing $\Phi_0$ by $\bar \Phi_0$

\bb\nn
Q(\theta)&=&
\int\frac{d\lambda}{2\pi}\exp \left [ i\lambda
\left \{
\theta
+\frac{1}{\sqrt{\tk_2}}
\frac{\partial}{\partial h_0}
\Phi_0(z_0,h_0)
\right \}
\right .
\\ \label{Qtheta}
&+&\left .\Phi_0(z_0-i\frac{\lambda z_0'}{\sqrt{\tk_2}},h_0-i\frac{\lambda}{\sqrt{\tk_2}})
-\Phi_0(z_0,h_0)
\right ].
\ee

This result is convenient since the Fourier transform depends
functionally only on $\Phi_0$ which can be written for different lattice
geometries or coupling distributions between the nearest-neighbor spins, by only
modifying the structure of the Green function $G_{\bq}$ for example. Also only
$z_0$ and $z'_0$ (or $\bar
z_0$ and $\bar z'_0$) are necessary for the evaluation of this expression by
solving \eref{solz0} and \eref{z0p} (or \eref{solz0b} and \eref{z0pb}). The
distribution is centered around the value

\begin{figure}
\centering
\includegraphics[scale=0.5,clip]{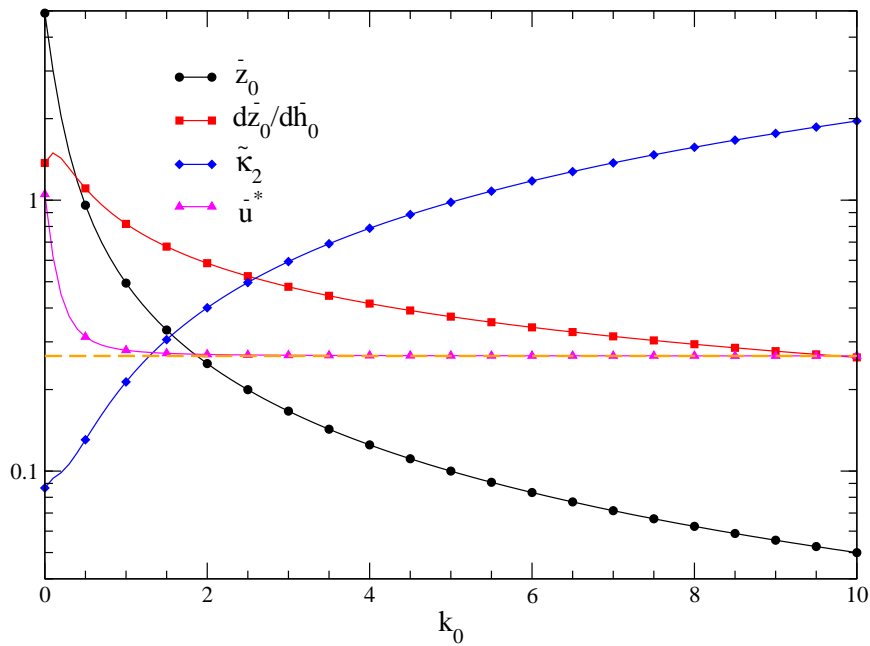}
\caption{\label{figZ} (Color online) plot of $\bar z_0$, $\bar z'_0$, $\tk_2$,
and asymptotic
slope $\bar u^*=\sqrt{\tk_2}\bar z_0/\bar z'_0$ as function of $k_0$ for the
distribution of $|m|$. In the limit of large $k_0$, $\bar u^*$ approaches the
constant $\bar u^*_{\infty}=\sqrt{\pi(1-3/\pi)/2}\simeq 0.266$ (dashed orange
line, see text).}
\end{figure}

\bb
c_0:=\frac{1}{\sqrt{\tk_2}}
\frac{\partial}{\partial h_0}\Phi_0(z_0,h_0)
\ee

which is equal to $c_0=h_0/(2\sqrt{\tk_2}K_cz_0)$ using \eref{phi0} or
$c_0=1/\sqrt{\pi \tk_2K_c\bar z_0}$ using \eref{phi0b} when $\bar h_0=0$ for
simplification. Typical behavior of quantities $\bar z_0$, $\bar z'_0$, and
$\tk_2$ are plotted in \efig{figZ} after solving \eref{solz0b} recursively.

\section{Asymptotic limits and numerical results}

The asymptotic analysis of \eref{Qtheta} when $\theta \gg 1$ or $\theta \ll -1$
is useful to determinate the behavior of the PDF for large deviations and
obtain its universal characteristics. The standard
method is to derive the argument in the exponential of the integral
\eref{Qtheta} with respect with
$\lambda$ and look for the dominant saddle point in the complex plane.
This is equivalent to analyze the existence of solutions of the following
equation:

\bb\nn
\theta+\frac{1}{\sqrt{\tk_2}}
\frac{\partial}{\partial h_0}
\Phi_0(z_0,h_0)&=&
\frac{z_0'}{\sqrt{\tk_2}}
\frac{
\partial\Phi_0}{\partial z_0}(z_0-i\frac{\lambda
z_0'}{\sqrt{\tk_2}},h_0-i\frac{\lambda}{\sqrt{\tk_2}})
\\ \label{asympt}
&+&
\frac{1}{\sqrt{\tk_2}}
\frac{\partial\Phi_0}{\partial h_0}(z_0-i\frac{\lambda
z_0'}{\sqrt{\tk_2}},h_0-i\frac{\lambda}{\sqrt{\tk_2}}).
\ee

The detailed analysis of this equation is given in \ref{app3}. We
find two kinds of behavior.
For $\theta\ll -1$ the PDF is exponentially decreasing $Q(\theta)\simeq \exp(u^*\theta)$ with
coefficient $u^*$ (and $\bar u^*$) equal to $u^*:=\sqrt{\tk_2}z_0/z'_0>0$
($\bar u^*:=\sqrt{\tk_2}\bar z_0/\bar z'_0>0$). The saddle point
structure can be checked numerically by fitting the asymptotic function
given below by \eref{QS} where corrections have to be taken into
account for moderate $\theta<0$ in addition to the dominant exponential term.
In particular three dominant contributions are present corresponding
respectively to the linear behavior discussed above, plus a term in
$\sqrt{|\theta+c_0|}$, and a logarithm $\log|\theta+c_0|$:

\bb\label{QS}\fl
Q(\theta)
\simeq
\exp\left (
u^*\theta+
\frac{\sqrt{u^*}|h_0-u^*/\sqrt{\tk_2}|}{2\sqrt{K_cz_0}}\sqrt{|\theta+c_0|}
-\ff\log |\theta+c_0|
\right ),\;\;\theta\ll -1.
\ee

In the opposite limit, when $\theta\gg 1$, the PDF falls off
exponentially with a stretched exponent equal to $d/(d-2)=3$ in three
dimensions. The saddle point evaluation in this case is detailed in
\ref{app3}:

\bb\label{QL}
Q(\theta)\simeq \exp\Big (-\co(d) \,\left (
\frac{\sqrt{\tk_2}}{z'_0}\,\theta\right )^{d/(d-2)}\Big ),\;\;\theta \gg 1,
\ee

with eventual corrections. Here the expression of coefficient $\beta(d)$ is
given by \eref{beta}, and is simply equal to $\beta(3)=8\pi^2$ in three
dimensions.
These results can be checked numerically after plotting the Fourier integral
\eref{Qtheta}.

\begin{figure}
\centering
\includegraphics[scale=0.45,clip]{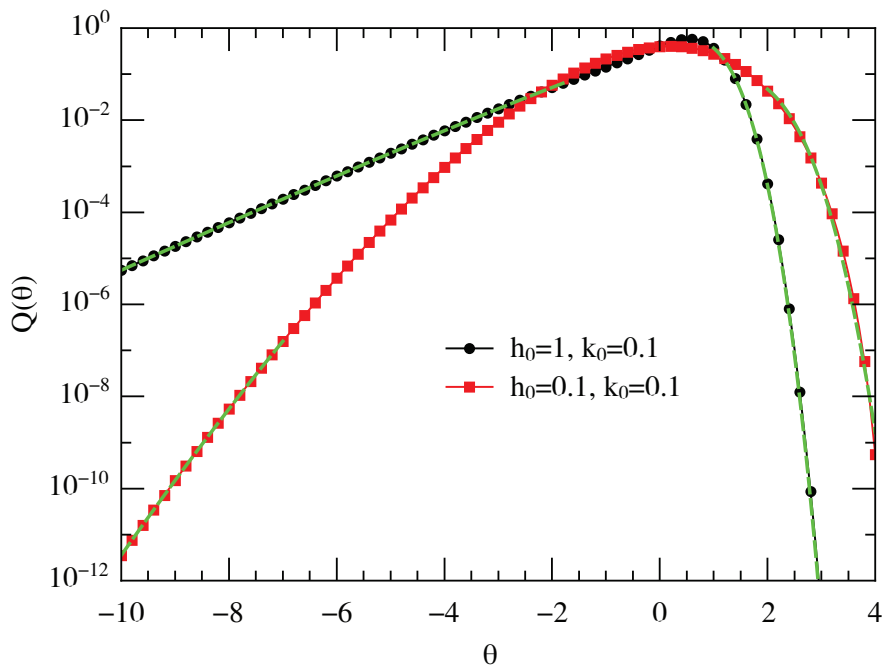}
\caption{\label{figQh} (Color online) plot of distribution $Q(\theta)$
\eref{Qtheta} for $m$ and for two different sets of parameters
$(h_0,k_0)$ ($L=10$). Dashed lines correspond to approximation curves given by
asymptotic regime detailed in \ref{app3}. For negative deviations, the
behavior is found to be exponential \eref{QS} with $u^*\simeq 7.396$ for
$(h_0=0.1,k_0=0.1)$ and $u^*\simeq 1.44$ for $(h_0=1,k_0=0.1)$, close to
expected values $u^*=8.15$ and $u^*=1.45$ respectively. For positive deviations
$\theta >1$, curves are fitted with a stretched exponential with cubic exponent
\eref{QL}.}
\end{figure}

In \efig{figQh} is represented the distribution for $m$ and for two different
sets of parameters $(h_0,k_0)$. $Q(\theta)$ is
typically non-Gaussian, with a exponential behavior for negative deviations, as
expected, and more pronounced falloff for positive deviations for which a
stretched exponential with a cubic exponent was performed adequately. The
exponential behavior is typical to
extreme value statistics when one studies the extrema distribution of a
set of independent random variables, such as the Gumbel distribution.
This distribution has a more pronounced asymptotic double-exponential
falloff with a coefficient which is integer. Here the coefficient of $\theta^3$
is non-integer and depends on the saddle point structure. When field $h_0$
is increased, negative deviations are enhanced, whereas the curve falls off more
rapidly for $\theta>0$ as the coefficient $(\sqrt{\tk_2}/z'_0)^3$ in
\eref{QL} increases.

\begin{figure}
\centering
\includegraphics[scale=0.45,clip]{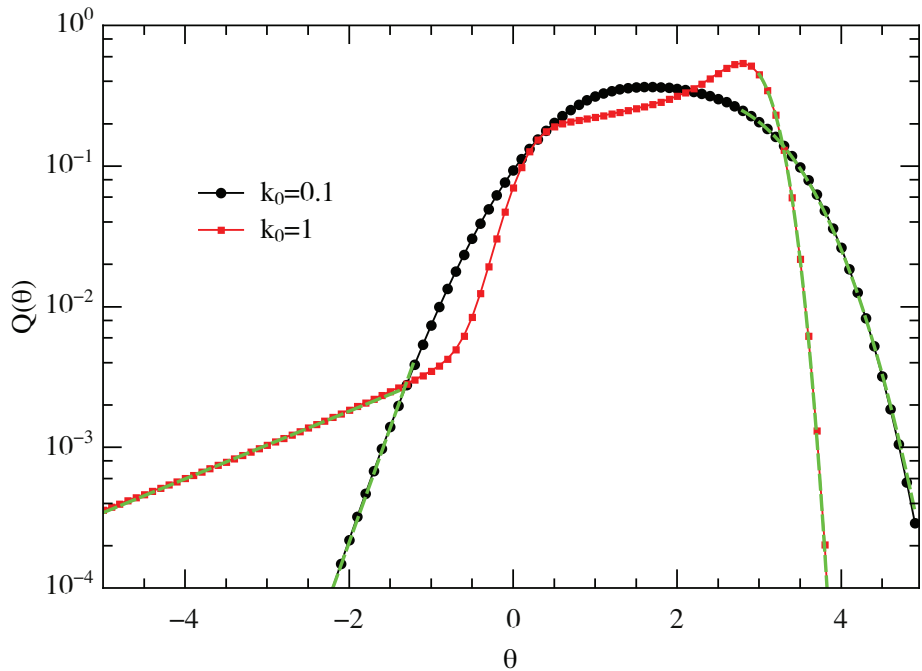}
\caption{\label{figQk} (Color online) plot of distribution $Q(\theta)$
\eref{Qtheta} for $|m|$ and
for two different parameters $k_0$ ($\bar h_0=0$ and $L=10$). Dashed lines
correspond to approximation curves given by the asymptotic regime. For negative
deviations, the fits are
given by the exponential behavior \eref{QS} with $\bar u^*\simeq 0.26$ for
$k_0=1$, close to the expected value $\bar u^*=0.279$. For $k_0=0.1$, the
asymptotic regime is not reached since we expect a slope $\bar u^*=0.62$, lower
than the slope found in the interval $[-3,-1]$. For
positive deviations $\theta >1$, curves are fitted accurately with a stretched
exponential with cubic exponent \eref{QL}.}
\end{figure}

\begin{figure}
\centering
\includegraphics[scale=0.5,angle=0]{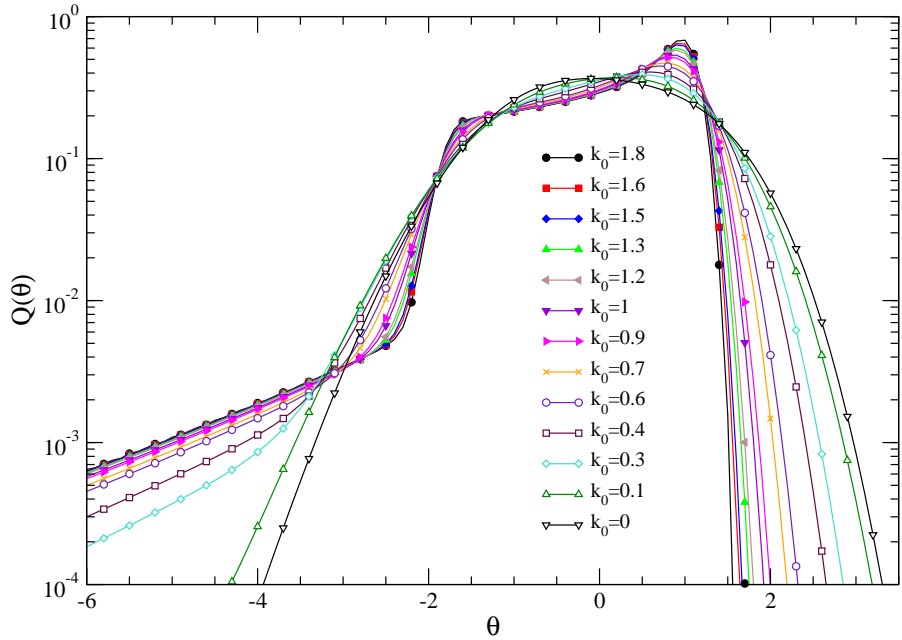}
\caption{\label{figQkM} (Color online) multiple plots of the distribution
$Q(\theta)$ for $|m|$ and for several values of $k_0$ ($\bar h_0=0$ and $L=10$).
A crossover occurs around $k_0=0.5$ below which the plateau-like feature of the
distribution is smoothed out. The large $k_0$ limit is almost reached when $k_0$
is larger than 2 (see also \efig{figQkL}).}
\end{figure}

 The distribution for $|m|$ follows the same trend. For large and negative
deviations, the saddle point solution is given by \eref{spb}, with an
asymptotic behavior also corresponding to \eref{QS} and \eref{QL}. \efig{figQk}
represents typical examples for $k_0=0.1$ and $k_0=1$ at zero field $\bar
h_0=0$.
The PDF is clearly non-Gaussian, and, for deviations left to the most probable
value, two distinct behaviors separated by a crossover interval. When
$\theta$ is
largely negative, the slope of the exponential asymptote is given by $\bar u^*$,
whereas there exists a plateau-like regime where the distribution, at least for
$k_0>1/2$, takes significant values in a whole range of $\theta$ values.
However, when $k_0$ decreases, the asymptotic regime is reached only for
negative deviations that become larger and larger, as it is shown in the set of
curves \efig{figQkM} where we have varied $k_0$ almost continuously.

In particular, in this figure, the limit $k_0\gg 1$ is seen to be reached
for $k_0$ values close to 2, and the limiting distribution takes a universal
form.

Indeed the saddle point values when $\bar h_0=0$ can be computed exactly in this
limit since $\bar z_0\sim 1/(2k_0)$, $\bar z'_0\sim\sqrt{\bar z_0}$, and
$\tk_2\sim 1/\bar z_0$,
and $\bar u^*$ approaches the numerical constant $\bar
u^*_{\infty}=\sqrt{\pi(1-3/\pi)/2}\simeq 0.266$, see \ref{app3},
\eref{univ}. We find that the limiting value $Q_{\infty}$ of
the distribution is given by

\bb\nn
Q_{\infty}(\theta)&=&\int \frac{\bar u^*_{\infty} d\,\lambda}{2\pi}
\exp\left \{ i\lambda\Big (\bar u^*_{\infty}\theta+\frac{1}{\sqrt{\pi}}\Big )
-\ff\log \Big (1-i\lambda\Big )\right .
\\ \label{Qlim}
&-&\left .\frac{\pi\lambda^2}{4(1-i\lambda)}
+\log\Big [1-\erf\Big (\frac{i\sqrt{\pi}\lambda}{2\sqrt{1-i\lambda}}\Big )\Big ]
\right \}.
\ee

\begin{figure}
\centering
\includegraphics[scale=0.5,angle=270]{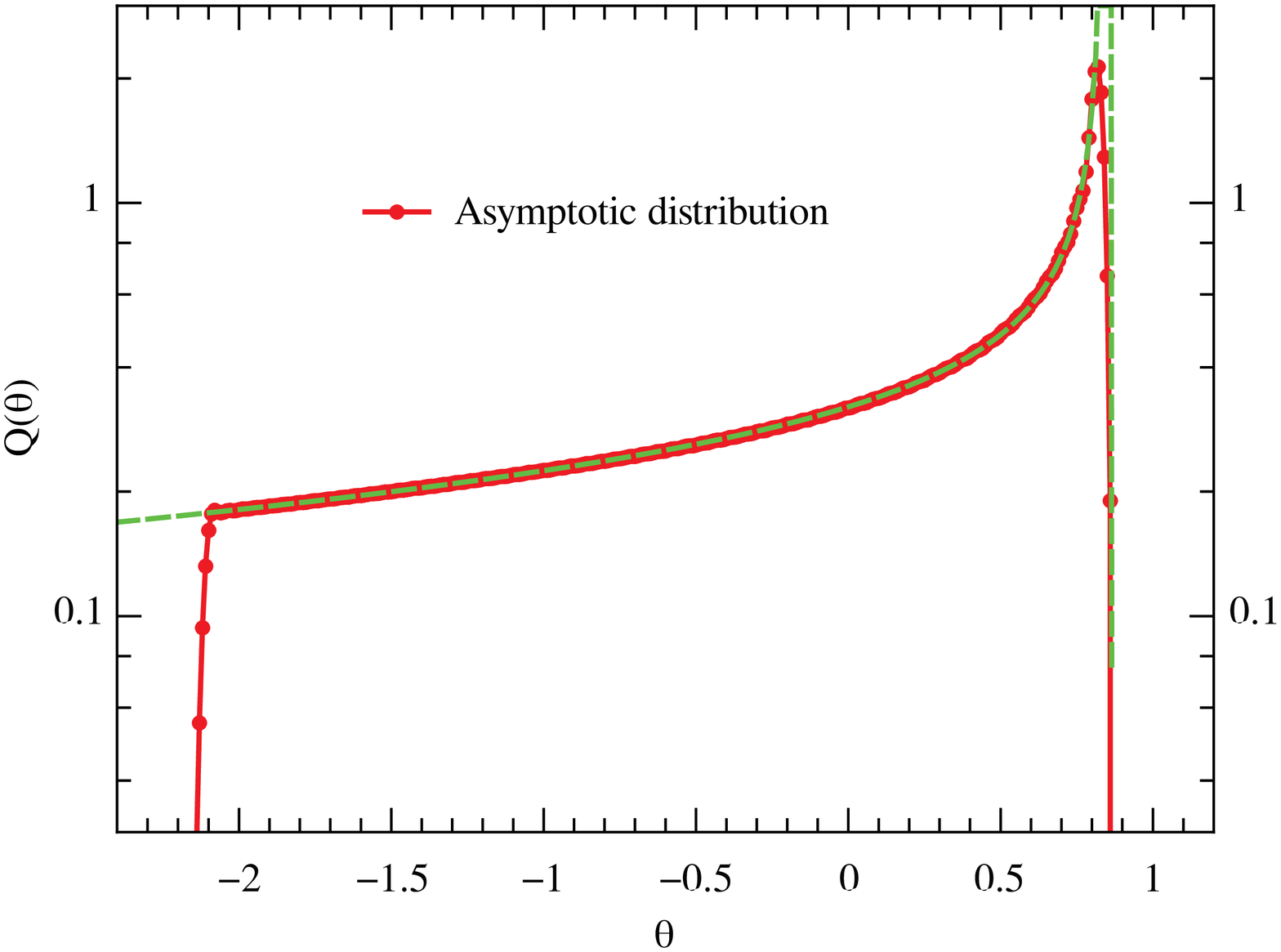}
\caption{\label{figQkL} (Color online) plot of the distribution
$Q_{\infty}(\theta)$ \eref{Qlim} in the limit of large $k_0$ ($\bar
h_0=0$ and $L=10$).
Dashed blue line corresponds to the approximate function
$Q_{\infty}(\theta)\approx
\exp\{a\theta-b/(\theta-\theta_c)\}/(\theta_c-\theta)^{\alpha}$ in
the interval $\theta<\theta_c$, with parameters $a=0.055$, $b=0.0351$,
$\theta_c=0.8533$,
and $\alpha=0.826$.}
\end{figure}

This integral, independent of $k_0$, is a universal function of $\theta$
with numerical factors only. It is plotted in \efig{figQkL} and presents a
dominant contribution in the interval $-2.1<\theta<0.9$, with a sharp decreasing
behavior outside this interval. The plot suggests that there may exist a
cutt-off at $\theta=\theta_c\simeq 0.86$ above which the distribution vanishes.
We have approximated quite accurately this function by the Ansatz
$Q_{\infty}(\theta)\approx
\exp\{a\theta-b/(\theta-\theta_c)\}/(\theta_c-\theta)^{\alpha}$ in
the interval $\theta>-2$, with fitted parameters $a=0.055$, $b=0.0351$,
$\theta_c=0.8533$ and $\alpha=0.826$.

\begin{figure}
\centering
\includegraphics[scale=0.5,angle=0]{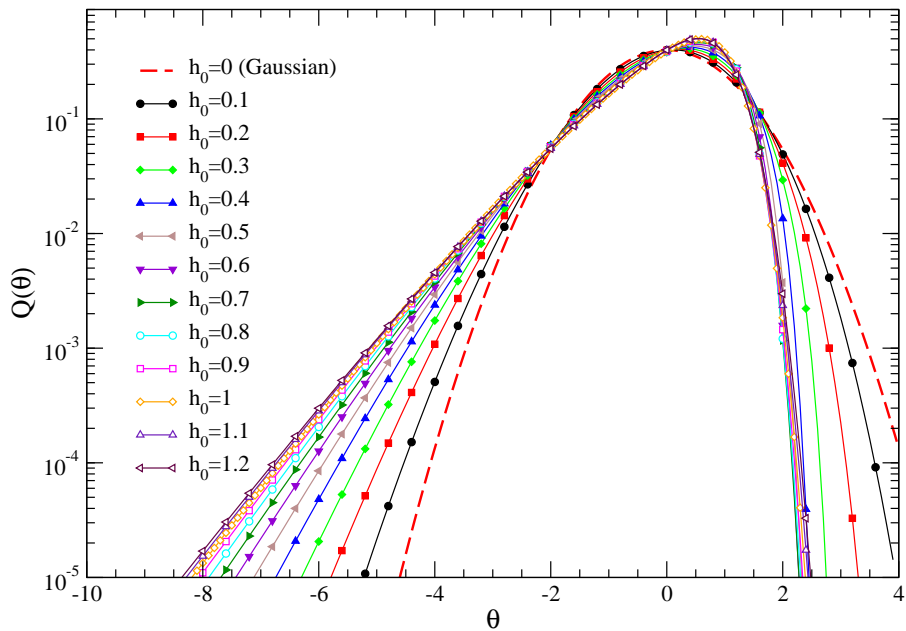}
\caption{\label{figQhM} (Color online) multiple plots of the $m$ distribution
$Q(\theta)$ for several values of $h_0$ ( $k_0=0$ and $L=10$).
The distribution at $h_0=0$ is purely Gaussian
$Q(\theta)=\exp(-\theta^2/2)/\sqrt{2\pi}$ (see text).
}
\end{figure}

 Finally, in \efig{figQhM}, we have plotted the distribution of $m$ for a series
of several $h_0$ values at $k_0=0$. As discussed above, the distribution for
($h_0=0,k_0=0$) is a Gaussian, and differs from the Gaussian form as
the field is increased, and presents asymptotically the exponential form
\eref{QS} for large negative deviations. However the structure is different
from \efig{figQkM} since no plateau regions are present.

\section{Conclusion}

In this paper, we presented an analytical method to compute the distribution
function of the magnetization in the spherical model at criticality, with
a numerical application in three dimensions. The advantage of this
spin-correlated model is that the critical region can be explored analytically
leading to important properties of the PDF. In particular, the
exponential behavior for large and negative
deviations are characteristic of the presence of soft modes $|\bq|\ll 1$
destroying the long-range order as seen in the 2D-XY model, with additional
corrections not seen usually in extreme statistics distributions of uncorrelated
variables.
For larger deviations, the PDF is falling off more rapidly than a Gaussian, as
the long-range order is more difficult to restore. The stretched exponent of the
exponential falloff depends on the dimension like $d/(d-2)$. When $d=4$ the
Gaussian behavior is recovered, and when $d$ is close to two dimensions, the
exponent is
diverging, and we expect formally a double exponential falloff in this limit for
positive deviations, as for the 2D-XY model and 2D Ising model as well
~\cite{epl06}. By inspection of the partition function in the Fourier space
\eref{ZNq}, we can reformulate the problem and associate the PDF as the
distribution of a set of uncorrelated Fourier modes $S_{\bq}$ which appear in
the integrand. However these
modes are coupled by the variable of integration $s$ coming from the constraint
over the spins, which leads, a priori, to a PDF which is not a standard
distribution. It seems nonetheless that using a saddle point analysis in the
large size limit we obtain the PDF of quasi-independent modes not evenly
distributed, which implies strongly non-Gaussian characteristics in the general
case~\cite{bertcl08}. We found for the spherical model that the PDF belongs to a
class of functions closely related to Gumbel but with a stretched exponential
for positive deviations instead of a double exponential form when the dimension
is larger than two. Moreover the corrective terms found asymptotically for
negative deviations in \eref{QS} may originate from the correlations between
modes and induced by the constraint over the spins.

 The main result \eref{Qtheta}, giving formally the Fourier expression of the
PDF, can be applied for any model where a saddle point analysis is exact,
since the cumulants can be expressed simply with a generating function, see
\ref{app2}. This model can moreover be modified in order
to incorporate additional couplings or crystal fields for example. In this
case, only the rescaled free energies given by \eref{phi0} and \eref{phi0b} have
to be reformulated to accommodate the modifications made in the original
Hamiltonian.


\appendix

\section{\label{app1}Limiting distribution}

Cumulant $\kappa_p$ are defined by the characteristic function $H$ of
instantaneous spin order parameter $\sum_iS_i/N$

\bb\label{charH}
{\mathbb E}\left [ e^{i\lambda \sum_iS_i/N} \right ]=\exp [H(\lambda)]=
\exp\left (\sum_{p\geq 1}\kappa_p \frac{(i\lambda)^p}{p!}
\right )
\ee

where ${\mathbb E}[.]$ is the thermal average operator over the different spin
configurations and $\kappa_p$ the cumulants. The distribution $P(m,L)$ can then
be expressed as the Fourier transform of $\exp[H(-\lambda)]$

\bb\nn
P(m,L)&:=&\int \frac{d\lambda}{2\pi} \exp [i\lambda m+H(-\lambda)]
\\ \label{Pcum}
&=&\int \frac{d\lambda}{2\pi} \exp \left [
i\lambda m+\sum_{p\geq 1}\kappa_p \frac{(-i\lambda)^p}{p!}\right ].
\ee

If $\kappa_p=0$ for all $p>2$, $P(m,L)$ is Gaussian and $\kappa_2$
is the variance $\sigma^2$,
$P(m,L)=\exp \left [-\frac{(m-\kappa_1)^2}{2\kappa_2} \right
]/\sqrt{2\pi \kappa_2}$ or $P(m,L)=Q\left
(\theta:=\frac{(m-\kappa_1)}{\sqrt{\kappa_2}} \right )/\sqrt{\kappa_2}$, where
$Q(\theta)=\exp(-\theta^2/2)/\sqrt{2\pi}$ is the normalized Gaussian
distribution function. In
general, we expect the distribution to be non-Gaussian if $\kappa_{p\ge 3}\ne
0$, and $P(m,L)$ can be put generally into the following form using the rescaled
cumulants defined in \eref{kp}

\bb\nn\fl
P(m,L)&=&L^{\beta/\nu}\int\frac{d\lambda}{2\pi\sqrt{\tk_2}}\exp \left [
i\lambda\frac{(m-\kappa_1)}{\sqrt{\kappa_2}}
+\sum_{p\geq 2}\frac{\tk_p}{\tk_2^{p/2}} \frac{(-i\lambda)^p}{p!}\right ]
\\ \label{Plim}\fl
&=&
L^{\beta/\nu}\int\frac{d\lambda}{2\pi\sqrt{\tk_2}}\exp \left [ i\lambda\theta
+\sum_{p\geq 2}\frac{\tk_p}{\tk_2^{p/2}} \frac{(-i\lambda)^p}{p!}\right
]=:L^{\beta/\nu}\frac{1}{\sqrt{\tk_2}}Q(\theta),
\ee

with $\beta/\nu=(d-2)/2$ for the spherical model.


\section{\label{app2}Cumulant expression}

In this appendix, we derive the exact differential relation \eref{cumulant}.
Consider the $p^{th}$ cumulant of the distribution as the
derivative of the $(p-1)^{th}$ one, using the assumption that \eref{cumulant} is
correct for $\kappa_{p-1}$

\bb\fl
N^{p-1}\kappa_p&=&\frac{d}{dh}\left (
\sum_{k=0}^{p-1}\bin{p-1}{k}\frac{\partial^{p-1}\Phi}{\partial z^k\partial
h^{p-1-k}}
z'(h)^k \right )
\\ \nn\fl
&=&
\sum_{k=0}^{p-1}\bin{p-1}{k}
\left [
\frac{\partial^{p}\Phi}{\partial z^{k+1}\partial h^{p-1-k}}
z'(h)^{k+1}+
\frac{\partial^{p}\Phi}{\partial z^{k}\partial h^{p-k}}
z'(h)^{k}
+\frac{\partial^{p-1}\Phi}{\partial z^k\partial h^{p-1-k}}
kz'(h)^{k-1}z''(h)
\right ].
\ee

The first two terms in the last line can be brought together using the
binomial formula $\bin{p-1}{k-1}+\bin{p-1}{k}=\bin{p}{k}$  and rearranging the
summation index $k$. We then obtain

\bb\fl
N^{p-1}\kappa_p=
\sum_{k=0}^{p}\bin{p}{k}
\frac{\partial^{p}\Phi}{\partial z^{k}\partial h^{p-k}}
z'(h)^{k}+z''(h)
\sum_{k=0}^{p-1}k\bin{p-1}{k}
\frac{\partial^{p-1}\Phi}{\partial z^k\partial h^{p-1-k}}
 z'(h)^{k-1}.
\ee

Using $k\bin{p-1}{k}=(p-1)\bin{p-2}{k-1}$, we can show that the last term is
equal to

\bb\nn\fl
z''(h)
\sum_{k=1}^{p-1}k\bin{p-1}{k}
\frac{\partial^{p-1}\Phi}{\partial z^k\partial h^{p-1-k}}
 z'(h)^{k-1}&=&
 z''(h)
\sum_{k=1}^{p-1}(p-1)\bin{p-2}{k-1}
\frac{\partial^{p-1}\Phi}{\partial z^k\partial h^{p-1-k}}
 z'(h)^{k-1}
 \\ \fl
 &=&z''(h)(p-1)\frac{d^{p-2}\partial_z\Phi}{dh^{p-2}}.
\ee

This term is equal to zero at the saddle point value, proving the recurrence
for \eref{cumulant} at order $p$. Now, if we consider the first expression
for $\Phi$ in \eref{phidef}, the dependence in $h$ is quadratic, and only 3
terms
remain from \eref{cumulant}

\bb
\kappa_p=\frac{1}{N^{p-1}}\left [
\frac{\partial^p \Phi}{\partial z^p}z'^p+p\frac{\partial^p \Phi}{\partial
z^{p-1}\partial h}z'^{p-1}
+\ff p(p-1)\frac{\partial^p \Phi}{\partial z^{p-2}\partial h^2}z'^{p-2}
\right ].
\ee

Using for example the exact expression for $\Phi(z,h,0)$, we can compute the
cumulants as function of the saddle point $z$ and its derivative $z'(h)$ only

\bb\nn
\frac{(-1)^p}{p!}\kappa_p&=&\frac{z'^p}{2pN^pz^p}+\frac{z'^{p-2}}{4KN^{p-1}z^{
p-1}}
+\frac{z'^p}{2p}\frac{1}{N^p}\sum_{\bq\ne 0}\frac{1}{\left
(z+d-\sum_{\alpha}\cos q_{\alpha}\right )^p}
\\ \label{kappa}
&-&\frac{hz'^{p-1}}{2KN^{p-1}z^{p}}+\frac{h^2z'^{p}}{4KN^{p-1}z^{p+1}}.
\ee

This result can be used in \eref{Pcum} to obtain the characteristic function
$H(-\lambda)$ by re-summation over $p$ of the cumulant series.


\section{\label{app3}Asymptotic analysis}

The asymptotic solutions for the saddle point equation \eref{asympt} can be
derived when $\theta \ll-1$ and $\theta \gg 1$. In both cases, the left hand
side of \eref{asympt} proportional to $\theta$ is diverging negatively and positively respectively,
which leads us to look for a diverging solution for the right hand side as
well. For the two cases considered in this paper, \eref{asympt} is equal to

\bb\nn
\theta+\frac{1}{\sqrt{\tk_2}}\frac{h_0}{2K_cz_0}
&=&
\frac{z_0'}{\sqrt{\tk_2}}
\left [
-\frac{1}{2Z_0}+\ff\sum_{\bq\ne 0}\frac{G^2_{\bq}}{L^4}\frac{Z_0}{1+
Z_0G_{\bq}/L^2}-\frac{H_0^2}{4K_cZ_0^2}+k_0
\right ]
\\ \label{asym}
&+&\frac{1}{\sqrt{\tk_2}}\frac{H_0}{2K_cZ_0},
\ee

and

\bb\fl\label{asymb}
\theta&+&
\frac{1}{\sqrt{\tk_2\pi K_c\bar z_0}}
=
\frac{\bar z_0'}{\sqrt{\tk_2}}
\left [
-\frac{1}{2\bar Z_0}+\ff\sum_{\bq\ne 0}\frac{G^2_{\bq}}{L^4}\frac{\bar
Z_0}{1+\bar Z_0G_{\bq}/L^2}-\frac{\bar H_0^2}{4K_c\bar Z_0^2}+k_0
\right .
\\ \fl \nn
&-&\left .\frac{\bar H_0}{2\sqrt{\pi K_c} \bar Z_0^{3/2}}\frac{\e^{-\bar
H_0^2/(4K_c\bar
Z_0)}}{1+\erf\Big (\frac{\bar H_0}{2\sqrt{K_c\bar Z_0}} \Big )}
\right ]
+
\frac{1}{\sqrt{\tk_2}}
\left [
\frac{\bar H_0}{2K_c\bar Z_0}
+\frac{1}{\sqrt{\pi K_c \bar Z_0}}\frac{\e^{-\bar H_0^2/(4K_c\bar
Z_0)}}{1+\erf\Big (\frac{\bar H_0}{2\sqrt{K_c\bar Z_0}} \Big )}
\right ],
\ee

where

\bb\nn
Z_0&=&z_0-i\frac{\lambda z'_0}{\sqrt{\tk_2}},\;\;
\bar Z_0=\bar z_0-i\frac{\lambda \bar z'_0}{\sqrt{\tk_2}}
\\
H_0&=&h_0-i\frac{\lambda}{\sqrt{\tk_2}},\;\;\bar H_0=-i\frac{\lambda}{\sqrt{\tk_2}}.
\ee

\begin{figure}
\centering
\includegraphics[scale=0.8]{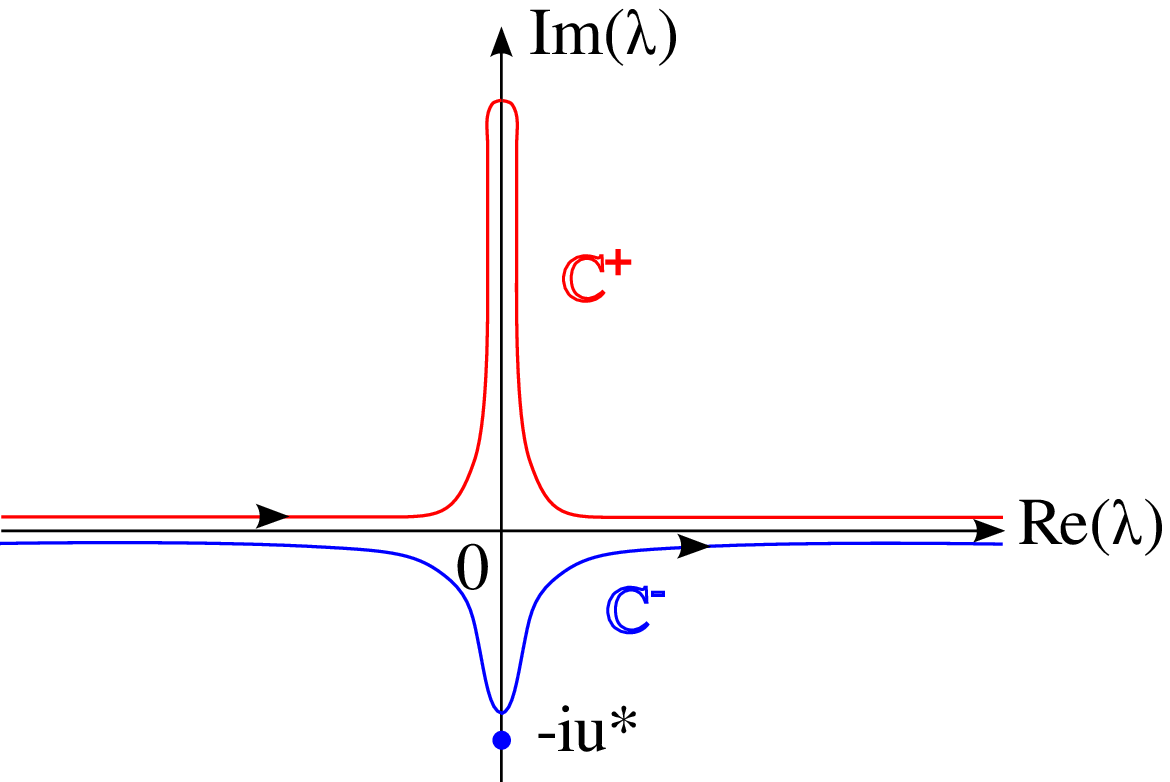}
\caption{\label{fig_asympt} (Color online) modified path of integration for the
saddle point analysis.
$C^+$ is the path chosen when $\theta\gg 1$ and $C^-$ when $\theta\ll -1$.}
\end{figure}

Considering first the regime $\theta\ll -1$, let assume in the following that $z_0$ and $z'_0$ are
positive (as well as $\bar z_0$ and $\bar z'_0 $), which will be realized
numerically by solving \eref{solz0} and \eref{z0p}. We are looking at diverging
terms in the previous saddle point equation coming from the inverse powers
of $Z_0\simeq 0$ or $\bar Z_0\simeq 0$ for example. Considering the integration
path $C^-$ in the
complex plane for $\lambda$ (see \efig{fig_asympt}), it is clear there is a special
point on the negative imaginary axis $\lambda=-iu^*$ with $u^*:=\sqrt{\tk_2}z_0/z'_0>0$ ($\bar
u^*:=\sqrt{\tk_2}\bar z_0/\bar z'_0>0$) for which $Z_0=0$ and the right hand side of
\eref{asym} (and \eref{asymb}) is singular.

The most singular term appears to be the one proportional to
$1/Z_0^2\propto 1/(u^*-u)^2$, after setting $\lambda=:-iu$ with $u$ close enough
to $u^*$, and which gives
the estimate for the saddle point solution

\bb\label{sp}
u\simeq u^*-\frac{\sqrt{u^*}|h_0-u^*/\sqrt{\tk_2}|}{2\sqrt{K_cz_0|\theta+c_0|}}.
\ee

Replacing this value in \eref{Qtheta} and integrating the Gaussian
fluctuations around the saddle point, we obtain the dominant contribution
for the large negative deviations \eref{QS}.

%

The asymptotic behavior of the PDF is therefore exponentially decreasing,
with a coefficient equal to $u^*$. The same
analysis for \eref{asymb} leads to a similar result, with coefficient $\bar u^*$
instead. The inverse powers in \eref{asym} and \eref{asymb} appear to be indeed
only in $1/Z_0$, $1/Z^{1/2}_0$, $1/Z^{3/2}_0$, and $1/Z^2_0$, with the latter
one being the dominant contribution. The additional two terms in \eref{asymb}
proportional to

\bb
\frac{\e^{-\bar H_0^2/(4K_c\bar
Z_0)}}{1+\erf\Big (\frac{\bar H_0}{2\sqrt{K_c\bar Z_0}} \Big )}
\simeq
\frac{\e^{-\bar u^{*3}/[4K_c\tk_2\bar
z_0(\bar u^*-u)]}}{1-\erf\Big (\bar u^{*3/2}/[2\sqrt{K_c\tk_2\bar
z_0(\bar u^*-u)}] \Big )}
\ee

are singular in the limit $u\rightarrow \bar u^*$ since both numerator and denominator vanish.
However this ratio can be evaluated using the asymptotic behavior of the error function for large and
negative argument. This ratio diverges like $1/\bar Z_0^{1/2}$ and elevates the power of the
factor $1/\bar Z_0^{3/2}$ in the second line in \eref{asymb}, giving rise to a dominant contribution proportional
to $1/\bar Z_0^{2}$, in addition to the identical contribution proportional to
$\bar H_0^2/\bar Z_0^2$ found in the first line.
Finally, the saddle point solution is given by

\bb\label{spb}
u\simeq \bar u^*-\frac{\bar u^*\sqrt{\bar u^{*}}}{\sqrt{2K_c\bar
z_0|\theta+c_0|}}.
\ee

instead of \eref{sp} and for $\bar h_0=0$ only. Since $Z_0$ is small, the sum
term
over modes $\bq$ in the saddle point equations is regular since it behaves in
this
limit like $Z_0\tg_2$ with $\tg_2$ finite. It is no more the case in $d\ge 4$
where $\tg_2$ is diverging logarithmically with the system size and where a
careful different analysis has to be made for the new saddle point, and where
one expects to find a Gaussian behavior with a $\theta^2$ contribution.

  Considering the PDF of $|m|$ in absence of field $\bar h_0=0$, we can
estimate the asymptotic value of $\bar u^*$ when $k_0$ is large. Indeed,
from \eref{solz0b} we obtain the approximation $\bar z_0\simeq 1/(2k_0)\ll 1$,
and from \eref{z0pb}, $\bar z'_0\simeq \sqrt{\bar z_0/(\pi K_c)}$. In addition,
the second cumulant is equal to $\tk_2\simeq (1-3/\pi)/(2K_c\bar z_0)$. Then we
obtain the universal limit

\bb\label{univ}
\lim_{k_0\rightarrow\infty}\bar
u^*=:\bar u^*_{\infty}=\sqrt{\frac{\pi}{2}\Big (1-\frac{3}{\pi}\Big )}\simeq
0.266.
\ee

In the regime of large and positive deviations $\theta\gg 1$, the previous
analysis can not be valid anymore because of the sign of $\theta$. Instead we
are looking for a path $C^+$ in the upper complex plane $\Im (\lambda)>0$ where
$Z_0$ is real and large. Indeed since $\theta$ is large, we can
assume that the modulus of $\lambda=:iu$ becomes large as well, so that
$Z_0\simeq uz'_0/\sqrt{\tk_2}\gg 1$. In this
case, terms proportional to inverse powers of $Z_0$ in \eref{asym} and
\eref{asymb} are small or finite, except the sum over the $\bq$ modes. This
sum contributes like $\sum_{\bq\ne 0}G_{\bq}/L^2$
which diverges with the system size when $d\ge 2$. Indeed
$g_1=\sum_{\bq\ne 0}G_{\bq}/L^d$ is finite for $d>2$ and diverges logarithmically
when $d=2$. Therefore the former sum is divergent with $Z_0$. Since $Z_0>0$ when
$\lambda=iu$ with $u>0$, the sign of the function

\bb\label{F0}
F(Z_0)=\sum_{\bq\ne 0}\frac{G^2_{\bq}}{L^4}\frac{
Z_0}{1+Z_0G_{\bq}/L^2}
\ee

is positive and consistent with the sign of $\theta$ on the left hand side of
\eref{asym}, and the saddle point equations reduces
to  $\theta\simeq z'_0F(Z_0)/(2\sqrt{\tk_2})$. The divergent part of $F(Z_0)$
can be evaluated exactly in the continuous limit. Indeed, we can rewrite
\eref{F0} as an integral

\bb
F(Z_0)\simeq \frac{L^{d}S_d}{(2\pi)^d}\int_{\cst/L}^{\cst} dq\,q^{d-1}
\frac{4}{q^4L^4}\frac{Z_0}{1+2Z_0/(q^2L^2)},
\ee

where $S_d=2\pi^{d/2}/\Gamma(d/2)$ is the hyper-spherical volume.
Performing the change of variable $qL/\sqrt{2Z_0}\rightarrow q$ and
taking afterward the limit $L\rightarrow\infty$, we obtain typically

\bb\fl
F(Z_0)\simeq
\frac{S_{d}2^{d/2}}{(2\pi)^d}\int_{\cst/\sqrt{2Z_0}}^{\infty}dq\,
\frac{q^{d-3}}{1+q^2}Z_0^{(d-2)/2}\simeq
\frac{\pi Z_0^{(d-2)/2}}{(2\pi)^{d/2}
\Gamma(d/2)\cos[\pi(d+1)/2]},
\ee

where the integral can be computed in general dimension $2<d<4$ after
substituting the lower bound with zero. Then the saddle point is given by

\bb\nn\fl
u&\simeq& \Big \{(2/\pi)(2\pi)^{d/2}\Gamma(d/2)\cos[\pi(d+1)/2]\Big \}^{2/(d-2)}
\left ( \frac{\sqrt{\tk_2}}{z'_0} \right )^{d/(d-2)}
\theta^{2/(d-2)}\gg 1
\\ \label{beta}\fl &=:& \co(d)\,\left ( \frac{\sqrt{\tk_2}}{z'_0}\right
)^{d/(d-2)}
\theta^{2/(d-2)}.
\ee

After inserting this value in \eref{Qtheta}, we finally obtain the dominant
behavior of the PDF given by the stretched exponential \eref{QL}.

%


\section*{References}
\bibliographystyle{iopart-num}
\bibliography{biblio_distrib}

\end{document}